\documentclass[twocolumn,preprintnumbers,amsmath,amssymb]{revtex4}
\usepackage{graphicx}
\usepackage{dcolumn}
\usepackage{bm}

\usepackage{graphicx}  
\begin{document}


\title{
Effect of plasma resonances on dynamic characteristics of  double graphene-layer optical modulator
}
\author{V.~Ryzhii\footnote{Electronic mail: v-ryzhii(at)riec.tohoku.ac.jp}$^{1,3,5}$,
T.~Otsuji$^{1,3}$
M.~Ryzhii$^{2,3}$, V.~G.~Leiman$^4$, S.~O.~Yurchenko$^5$, V.~Mitin$^6$, 
and  M.~S.~Shur$^7$
}
\affiliation{
$^1$Research Institute for Electrical Communication.Tohoku University, Sendai 980-8577, Japan\\
$^2$Computational Nanoelectronics Laboratory, University of Aizu,Aizu-Wakamatsu  965-8580, Japan\\
$^{3}$Japan Science and Technology Agency, CREST, Tokyo 107-0075, Japan\\
$^4$ Department of General Physics, Moscow Institute of Physics and Technology,
Dolgoprudny, Moscow Region 141700, Russia\\
$^5$Center for Photonics and Infrared Engineering,  Bauman Moscow State Technical University,
Moscow 105005, Russia\\
$^6$Department of Electrical Engineering,
University at Buffalo, State University of New York, NY 14260, USA\\ 
$^{7}$Department of   Electrical, Electronics, and Systems Engineering, Rensselaer 
Polytechnic Institute, 
Troy, NY 12180, USA\\
}

\begin{abstract}
We analyze the  dynamic operation of an  optical modulator based 
on  double graphene-layer(GL) structure utilizing
the variation of the GL absorption  due to the
electrically controlled Pauli blocking effect. The developed device model yields
the dependences of the modulation depth on 
the control voltage and the modulation frequency.
The excitation of plasma oscillations in double-GL structure can result in the 
resonant increase of the modulation depth, when the modulation frequency approaches the plasma
frequency, which corresponds to the terahertz frequency for the typical parameter values. 
\end{abstract}

\maketitle
\newpage

\section{Introduction}
The gapless 
energy spectrum of graphene layers (GLs))~\cite{1}, results in the interband absorption 
of the electromagnetic
radiation from the terahertz  to ultraviolet range\cite{2}.
This opens up prospects to use graphene structures in active and passive
optoelectronic devices.
Novel lasers, photodetectors, modulators, and mixers have been 
proposed and studied
~\cite{3,4,5,6,7,8,9,10,11,12,13,14,15} (see also the review paper~\cite{16} and references 
therein).
Varying the controlling voltage one can effectively change the electron 
and hole
densities and the Fermi energy and, hence, the intraband  and  interband absorption
in GLs.
 An increase in the electron (hole) density increases  the intraband absorption
 (associated with the Drude mechanism), but it decreases 
  the interband absorption
 due to  
the Pauli blocking effect. 
Which mechanism dominates depends on the incident 
photon energy $\hbar\Omega$
and the momentum relaxation time of electrons and holes, $\tau$.
At the frequencies of infrared and visible radiation, $\Omega \gg \tau^{-1}$, so that the Drude
absorption is weak. Even in  the terahertz (THz) range, the latter inequality can  be valid 
for high quality GLs (like those studied in Refs.~\cite{17,18}),
even at room temperatures. 

\begin{figure}[t]\label{Fig.1}
\begin{center}
\includegraphics[width=6.3cm]{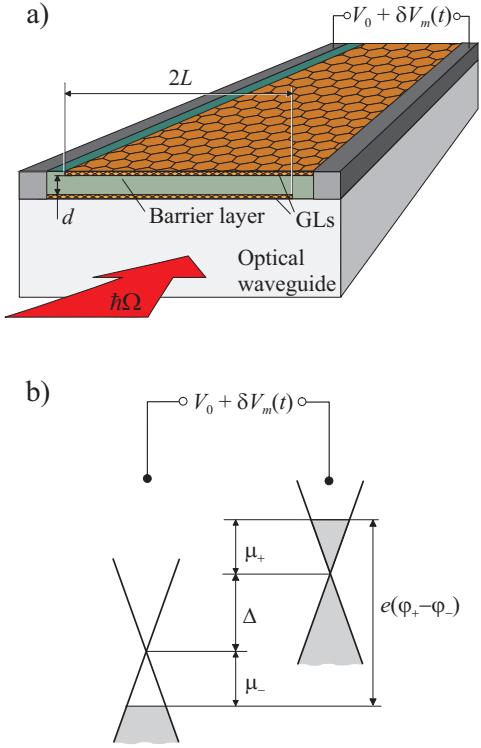}
\caption{
Schematic view of (a) double-GL modulator structure coupled with an optical waveguide and (b) band diagram of GLs
under a voltage drop between them (shaded areas indicate the states occupied by electrons)
.}
\end{center} 
\end{figure}

In this paper, we develop a device model of a double-GL optical modulator
proposed and demonstrated in Ref.~\cite{19}. In this modulator, GLs were separated by relatively thick barrier and the structure was integrated with an optical waveguide
The operation of the device under consideration is associated with the filling of GLs with electrons 
and holes injected from the contacts
under the self-consistent electric field created by the applied voltage
and the electron and hole charges in GLs. This process determines  both static and dynamic
characteristics of the double-GL modulator. At the non-stationary conditions, the dynamics
of the electron-hole plasma in double-GL can exhibit resonant response due to the excitation
of plasma oscillations similar to those well known in more traditional two-dimensional electron
and hole systems~\cite{20,21,22,23,24,25,26,27,28,29,30}. The plasma oscillations 
 in GL-structures also were considered previously
(see, for instance, Refs.~\cite{31,32,33,34,35}). A resonant plasmonic THz using
double-GL structures
was recently   studied in Ref.~\cite{36}.

In this paper, we develop a device model
for double-GL modulators of optical radiation
and demonstrate that the resonant
excitation of plasma oscillations in
double-GL modulators can provide an efficient modulation of optical radiation by
high frequency signals, in particular, in the THz range.

The variation of the Fermi energy in double-GL is associated with the electron and
hole injection and extraction by
the side contacts.
Therefore,
the consideration of the electron-hole plasma dynamics
in double-GLs must account for  the self-consistent
electric field  found from the solution of  the hydrodynamic equations and the Poisson equation.

\section{Equations of the model}

We consider the double-GL modulator reported in Ref.~\cite{19}. Its simplified structure  is shown in  Fig. 1(a). We assume that each GL is 
connected to one  side contact and is isolated from the opposite contact
(connected to the other GL),  The voltage, $V_m$, applied between these contacts, 
 $V_m = V_0 + \delta V_m$, where $V_0$ and 
$\delta V_m$ are the bias and modulation components.
We set
 $\delta V_m(t) = \delta V_m\exp(-i\omega\,t)$, where $\omega$ is the modulation frequency.
The latter is much smaller than the frequency of the incident optical radiation $\Omega$:
$\omega \ll \Omega$. The system of two highly conducting side contacts can be considered as a slot line
enabling the propagation of the modulation signals. The side contacts can also be connected to or be a part of  THz antenna,
which converts the incoming THz radiation into the modulation voltage.

The absorption coefficient of the light wave propagating along the waveguide with 
double-L on topGL is
determined by the real part of the double-GL conductivity 
${\rm Re}\sigma_{\Omega} = {\rm Re}\sigma_{\Omega}^{intra} + {\rm Re}\sigma_{\Omega}^{inter}$
at the frequency 
$\Omega$:

\begin{equation}\label{eq1}
\beta_{\Omega} =  \frac{4\pi{\rm Re}\langle\sigma_{\Omega}\rangle\Gamma_{\Omega}}
{c\sqrt{k}},
 \end{equation}
where $c$ is the speed of light in vacuum and $k$ is the dielectric constant of 
the waveguide material. The symbol $\langle ... \rangle$ means
the averaging  accounting for the distribution of the optical field
$E_{\Omega} (x,y)$ in the waveguide, where the $x$-axis and $y$-axis correspond  
to the direction
along double-GL structure and perpendicular to the $z$-axis corresponding to the
direction of the  wave propagation 
direction
in the waveguide. Thus,

\begin{equation}\label{eq2}
\langle\sigma_{\Omega}\rangle\Gamma_{\Omega} = 
\frac{\int_{-L}^{L} {\rm Re}\sigma_{\Omega}|E_{\Omega}(x,0)|^2dx}
{\int_{-\infty}^{\infty}\int_{-\infty}^{\infty}|E_{\Omega}(x,y)|^2dxdy}, 
\end{equation}
where
$$
\Gamma_{\Omega} = 
\frac{\int_{-L}^{L} |E_{\Omega}(x,0)|^2dx}
{\int_{-\infty}^{\infty}\int_{-\infty}^{\infty}|E_{\Omega}(x,y)|^2dxdy}.
$$
is the mode overlap factor and $2L$ is the GL length, which is approximately equal
to the spacing between the side contacts as shown in Fig. 1(a).

We assume that the electron and hole momentum relaxation time is associated with the 
scattering due to 
disorder and acoustic phonons, so that its energy dependence is given by 
$\tau^{-1} = \nu(\varepsilon/T_0)$, where $\nu$ is the characteristic scattering frequency
at $V = 0$ (at the Dirac point). In this case,
the real part of the  conductivity of two GLs at $\Omega \gg \nu_0$ can be presented as~\cite{10}

$$
{\rm Re}\sigma_{\Omega} 
= \biggl(\frac{e^2}{4\hbar}\biggr) 
\biggl\{2  
$$
$$
-\frac{1}{1 + \displaystyle\exp\biggl(\frac{\hbar\Omega/2 - \mu_{+}}{T}\biggr)}
 -  \frac{1}{1 + \displaystyle\exp\biggl(\frac{\hbar\Omega/2 + \mu_{+}}{T}\biggr)}
$$
$$
-  \frac{1}{1 + \displaystyle\exp\biggl(\frac{\hbar\Omega/2 - \mu_{-}}{T}\biggr)}
 -  \frac{1}{1 + \displaystyle\exp\biggl(\frac{\hbar\Omega/2 + \mu_{-}}{T}\biggr)}
$$
\begin{equation}\label{eq3}
+ \frac{8\nu}{\pi\hbar\Omega^2}\frac{(\mu_{+}^2 + \mu_{-}^2 + \pi^2T^2/3)}{T}\biggr\}. 
\end{equation}
Here 
$\mu_{+}$ and $\mu_{-}$ are the GL Fermi energies counted from the Dirac point in the upper and lower GLs, respectively, 
$T$ is the temperature, and
$e$ is the electron charge. 
The first five terms in Eq.~(4) correspond to the contributions of 
the interband transitions to the conductivity of the upper and lower GLs. These terms
explicitly account for the Pauli blocking effect. 
The last term in Eq.~(3) accounts for  the intraband transitions.
It is presented in the form providing an interpolated dependence of the 
intraband conductivity
on the Fermi energies and the temperature.
At slow variation of the applied voltage, $\mu_{+} = \mu_{-} = \mu$, where
$\mu$ obeys the following equation:
\begin{equation}\label{eq4}
2\mu + \Delta = eV_m
\end{equation}
The quantity $\Delta$ is determined by the electric field between GLs and the thickness of the barrier layer between GLs $d$.
Generally speaking (at sufficiently fast variations of the voltage), the spatial distributions of the electron and hole densities do not  follow the variation
of the applied voltage, and the Fermi energies depend 
on the coordinate $x$: $\mu_{+} = \mu_{+}(x)$ and  $\mu_{-} = \mu_{-}(x)$.

The Fermi energies in GLs are governed by the following equation: 
\begin{equation}\label{eq5}
|\Sigma_{\pm}| =  \frac{2}{\pi\hbar^2v_W^2}\int_0^{\infty}\frac{\varepsilon\,d\varepsilon}
{1 + \displaystyle\exp\biggl(\frac{\varepsilon - \mu_{\pm}}{T}\biggr)},
\end{equation}
where $v_W = 10^8$~cm/s is the characteristic velocity of electrons and holes in GLs.
Since the spacing, $d$, between GLs  is rather small compared to $2L$, one can use
the following formulas which relate the densities $\Sigma_{\pm}$ and 
the electric potentials
 of GLs $\varphi_{\pm}$:
 \begin{equation}\label{eq6}
\frac{4\pi\,e\Sigma_{\pm}}{k}=  \mp\frac{(\varphi_{+} - \varphi_{-})}{d}.
\end{equation}
The negative values of $\Sigma_{\pm}$ correspond to the case when a GL is filled by
electrons, whereas
the positive values correspond to the  filling by holes.

\section{Modulation characteristics}

When the control voltage $V_m$ varies slowly, one can set

\begin{equation}\label{eq7}
\varphi_{\pm} = \pm \frac{V_m}{2},
\end{equation}
and, hence,

\begin{equation}\label{eq8}
\Sigma_{\pm}=  \mp\frac{k\,V_m}{4\pi\,ed}, \qquad \mu_{\pm} = \mu,
\end{equation}
where $\mu$ is governed by the following equation:
\begin{equation}\label{eq9}
\frac{V_m}{\overline V} =  \int_0^{\infty}\frac{\xi\,d\xi}
{1 + \displaystyle\exp(\xi - \mu/T)}.
\end{equation}
Here 
\begin{equation}\label{eq10}
\overline V = \frac{8ed}{k}\biggl(\frac{T}{\hbar\,v_W}\biggr)^2.
\end{equation}
If $d = 10$~nm, $k = 7$ (Al$_2$O$_3$), and $T = 300$~K, from Eq.(10) 
we obtain $\overline V \simeq 30$~mV. 

At sufficiently large bias voltage $V_0$, the electron and hole systems in GLs become degenerate
(i.e., $\mu \gg T$),
and Eq.~(10) yields

\begin{equation}\label{eq11}
\mu \simeq T\sqrt{\frac{2V_m}{\overline V}} = \hbar\,v_W\sqrt{\frac{kV_m}{4ed}},
\end{equation}
so that, taking into account  Eq.~(4),
\begin{equation}\label{eq12}
 \Delta \simeq eV_m - 2T\sqrt{\frac{2V_m}{\overline V}} = eV_m - \hbar\,v_W\sqrt{\frac{kV_m}{ed}}.
\end{equation}

Using Eqs.~(1) and (3) and considering  Eq.~(8), we arrive at the following formula
for the absorption coefficient:

\begin{equation}\label{eq13}
\frac{\beta_{\Omega}}{ \overline\beta_{\Omega}} = 
1 -  
\frac{1}{1 + \displaystyle\exp\biggl(\frac{\hbar\Omega}{2T} - \sqrt{\frac{2V_m}{\overline{V}}}\biggr)}
+ \frac{32\nu\,T}{\pi\hbar\Omega^2}\biggl(\frac{V_m}{\overline V} + \frac{\pi^2}{12}\biggr). 
 \end{equation}
Here
\begin{equation}\label{eq14}
\overline \beta_{\Omega} =  \frac{2\pi\alpha \Gamma_{\Omega}}{\sqrt{k}}
\end{equation}
$\nu$ is the characteristic collision frequency of electrons and holes (which is
assumed to be proportional to $T$), and $\alpha = e^2/c\hbar \simeq 1/137$ is the fine structure constant, 
(so that $\pi\alpha \simeq 0.023$).


In the near infrared range of frequencies, $\hbar\Omega \gg T$,
 the contribution
of the intraband absorption at low control voltages $V$ is small. This implies that the
last term in the right-hand side of Eq.~(13) is much smaller than unity.

Considering a modulator for optical radiation with 
the wavelength $\lambda = 1537$~nm ($\hbar\Omega \simeq 0.8$)~eV as in Ref.~\cite{19},
setting $\hbar\Omega = 0.8$~eV, $V_0 = \overline{V}(\hbar\Omega/2\sqrt{2}T)^2
= (ed/k)(\Omega/v_W)^2 = 337$~mV, and $\nu = 10^{13}$~s$^{-1}$,
we obtain that the last term in Eq. (13) is about 0.03.
Thus, 
in the case of even modestly  perfect GLs, this term is relatively
small and can be omitted.  This implies that in such a case, the modulation is primarily due to the voltage control of the Pauli blocking but not due to the variations of the intraband absorption.  

In the most realistic case $\hbar\Omega \gg T$, the ratio of the intensities of
output (modulated by slow varying voltage)  and input  radiation, $I_0$  and $I_{00}$, respectively, taking into account that the second and fourth terms
yield the same contribution and omitting the third and fifth terms in Eq.~(14), can be presented as

\begin{equation}\label{eq15}
\frac{I_0}{I_{00}} = \exp\biggl[- \overline\beta_{\Omega}H  
\frac{\displaystyle\exp\biggl(\frac{\hbar\Omega}{2T} - \sqrt{\frac{2V_m}{\overline{V}}}\biggr)}
{1 + \displaystyle\exp\biggl(\frac{\hbar\Omega}{2T} - \sqrt{\frac{2V_m}{\overline{V}}}\biggl)}
\biggr]
\end{equation}
where $H$ is the double-GL length in the direction of radiation propagation.
and $\hbar\Omega_0 = \hbar\,v_W\sqrt{kV_0/ed} \propto \sqrt{V_0}$.
Equation ~(15) describes the variation of the output radiation $I_0$ caused by 
relatively slow variations of the applied voltage $V_m$ with arbitrary swing.
Using Eq.~(15), one can estimate the  modulation depth $m_0$ and the extinction ratio $\eta$ for the case 
of relatively slow modulation and when $V_m$ varies from
$V_m = 0$ to $V_m = V_m^{max} > \overline{V_m}^{max}$, where $\overline{V_m}^{max} = (4ed/k)(\Omega/v_W)^2$:
$m_0 = 1 - \exp(-\overline{\beta}_{\Omega}H)$ and $\eta = \exp(\overline{\beta}_{\Omega}H)$.

\section{Small signal linear modulation characteristics}

We now assume that $V_0$ is sufficiently large to form the degenerate electron and hole systems
in the pertinent GLs, while the time dependence of $\delta V_m(t) \ll V_0$ is still characterized 
by a small modulation 
frequency. In this case, from  Eq.~(15) we obtain
 the following formulas of the modulation amplitude $\delta I_0$ and the modulation depth 
 of the output radiation $\delta m_0 = \delta I_0/I$:
$$
\frac{\delta I_0}{I_{00}} = \overline\beta_{\Omega}H\exp\biggl\{- \overline\beta_{\Omega}H  
\frac{\displaystyle\exp\biggl[\frac{\hbar(\Omega - \Omega_0)}{2T}\biggr]}
{1 + \displaystyle\exp\biggl[\frac{\hbar(\Omega - \Omega_0)}{2T}\biggr]}\biggr\}
$$
\begin{equation}\label{eq16}
\times\frac{\displaystyle\exp\biggl[\frac{\hbar(\Omega - \Omega_0)}{2T}\biggr]}
{\biggl\{1 + \displaystyle\exp\biggl[\frac{\hbar(\Omega - \Omega_0)}{2T}\biggr]\biggr\}^2}\,\frac{\delta V_m}{\sqrt{2\overline{V}V_0}} .
 \end{equation}
 and 
\begin{equation}\label{eq17}
\delta m_0 = \overline\beta_{\Omega}H
\frac{\displaystyle\exp\biggl[\frac{\hbar(\Omega - \Omega_0)}{2T}\biggr]}
{\biggl\{1 + \displaystyle\exp\biggl[\frac{\hbar(\Omega - \Omega_0)}{2T}\biggr]\biggr\}^2}\,\,\frac{\delta V_m}{\sqrt{2\overline{V}V_0}} 
,
 \end{equation}
 where $\hbar\Omega_0 = \hbar\,v_W\sqrt{kV_0/ed} \propto \sqrt{V_0}$.
 
 In particular, at $eV_0 = \hbar\Omega$, when $\delta m_0$ reaches a maximum,
  Eqs.~(16) and (17) yield respectively
 
  \begin{equation}\label{eq18}
\frac{\delta I_0}{I_{00}} = \frac{\overline\beta_{\Omega}H}{4}
\exp\biggl(- \frac{\overline\beta_{\Omega}H}{2}  
\biggr)\,\frac{\delta V_m}{\sqrt{2\overline{V}V_0}} 
 \end{equation}

 \begin{equation}\label{eq19}
\delta m_0 = \frac{\overline\beta_{\Omega}H}{4}
\cdot\frac{\delta V_m}{\sqrt{2\overline{V}V_0}} 
 \end{equation}

\section{Plasma oscillations in double-GL structures}

\begin{figure}[t]\label{Fig.2}
\begin{center}
\includegraphics[width=6.3cm]{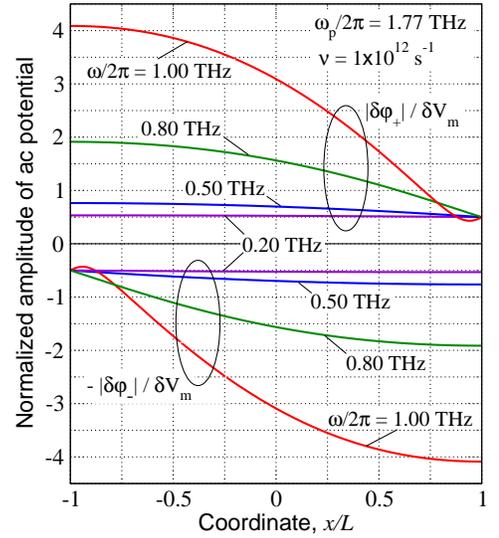}
\caption{Spatial distributions of the normalized amplitudes of the ac potential
$|\delta\varphi_+|/\delta V_m$ and $|\delta\varphi_-|/\delta V_m$ in
upper and lower GLs, respectively,  for different modulation frequencies $\omega/2\pi$. 
}
\end{center} 
\end{figure}

In the double-GL structures with sufficiently high conductivity 
at low modulation frequencies,
one can use Eq.~(6), i.e.,  put $\varphi_{\pm} = \pm V_m/2$.
However,  at elevated modulation frequencies
(for instance, in the THz range), the spatial distributions of the ac components of the
electron and hole charges, the electron and hole Fermi energies, and the self-consistent
electric potential are nonuniform because these distributions do not  follow  fast 
modulation signals. In sufficiently perfect GLs, the modulation signals can  excite
the electron-hole plasma oscillations.  
 In this situation, the ratio of the ac component of the radiation intensity
 $\delta I_{\omega}$ to the input intensity
 is given by the following equation, which replaces Eq.~(16):
$$
\frac{\delta I_{\omega}}{I_{00}} = \frac{\overline\beta_{\Omega}H}{4}\exp\biggl\{- \overline\beta_{\Omega}H  
\frac{\displaystyle\exp\biggl[\frac{\hbar(\Omega - \Omega_0)}{2T}\biggr]}
{1 + \displaystyle\exp\biggl[\frac{\hbar(\Omega - \Omega_0)}{2T}\biggr]}\biggr\}
$$
\begin{equation}\label{eq20}
\times\frac{\displaystyle\exp\biggl[\frac{\hbar(\Omega - \Omega_0)}{2T}\biggr]}
{\biggl\{1 + \displaystyle\exp\biggl[\frac{\hbar(\Omega - \Omega_0)}{2T}\biggr]\biggr\}^2}\,
\frac{1}{L}\int_{-L}^L\frac{dx\,(\delta \varphi_{+} - \delta\varphi_{-})}
{\sqrt{2\overline{V}V_0}}.
 \end{equation}
The last factor in the right-hand side of Eq.~(20) accounts for  the contributions of different parts of GLs being dependent on local values of the Fermi
energies and, hence, the local values of the ac potentials [see Fig.~1(b)]. 
 
To find the distributions of $\delta \varphi_{+}$ and $\delta \varphi_{-}$, one can use 
a system of hydrodynamic equations
(Euler equation and continuity equation) adjusted to the features 
of the electron and hole spectra in GLs~\cite{35}) for the electron and hole plasmas in 
the upper and lower GLs 
coupled with the Poisson equation. 
For simplicity, we use the Poisson equation in the gradual-channel approximation, 
which leads to Eq.~(6) above.
The linearized versions of the equations in question can be reduced 
to the following equation for the ac component of the potential at  the frequency $\omega$~\cite{36}:

\begin{equation}\label{eq21}
\frac{d^2\delta\varphi_{+}}{dx^2} 
+\frac{\omega(\omega + i\nu)}{s^2} (\delta\varphi_{+} - \delta\varphi_{-})= 0, 
\end{equation}
\begin{equation}\label{eq22}
\frac{d^2\delta\varphi_{-}}{dx^2} 
+\frac{\omega(\omega + i\nu)}{s^2} (\delta\varphi_{-} - \delta\varphi_{+})= 0. 
\end{equation}

Here $\nu \sim \nu_0$
is the collision frequency of electrons in GLs with impurities and acoustic phonons and $s$
is the characteristic velocity of plasma waves in GLs.
Since electrons and holes belong to different GLs separated by the rather high and thick barrier,
 their mutual collisions can be neglected.
The plasma-wave  velocity is determined by the net dc electron and hole density (i.e., by the Fermi energy)
$\Sigma_0 \simeq (\mu/\hbar\,v_W)^2/\pi$ 
and the gate layer thickness $d$~\cite{31,33,35}: $s \simeq v_W\sqrt{4\alpha_{GL} \mu_0d/\hbar\,v_W} \propto V_0^{1/4}d^{1/4}$, 
where $\alpha_{GL} = e^2/k\hbar\,v_W$
is the coupling constant ($\alpha_{GL}/\alpha = c/v_W \simeq 300$).
Similar equations were obtained for the gated electron channels in the traditional
heterostructures. The  difference is in the existence of two interacting 
channel and in  different values of $s$ . 
At $d = 10$~nm and $\hbar\Omega_0 = 0.2 - 0.8$~eV, one obtains $s/v_W \simeq 3.75 - 7.5$.

The plasma wave velocity in GL system 
is much higher than that in the standard gated channels~\cite{34,35}. 
Large  values of $s$ in the double-GL should allow us  
to achieve rather high plasma frequencies (in the THz range) 
in the  devices with relatively large values of $L$ (in the micrometer range).
If  the left-side contact and
the upper GL and the right-side contact and the lower GL, see Fig. 1(a)] are are very close~\cite{32},
one can disregard the gaps between GLs and the contacts and use  the following boundary conditions for Eqs.~(21) and (22):
\begin{equation}\label{eq23}
\delta\varphi_{+} |_{x = L}  = \frac{\delta V_m}{2}\exp(- i\omega t),\qquad
\delta\varphi_{-} |_{x = -L}  = -\frac{\delta V_m}{2}\exp(- i\omega t),
\end{equation}
\begin{equation}\label{eq24}
\frac{d \delta\varphi_{+}}{dx} \biggl|_{x = - L}  = 0, \qquad
\frac{d \delta\varphi_{-}}{dx} \biggl|_{x =  L}  = 0.
\end{equation}
 The latter boundary condition reflects the fact that the electron and hole 
 currents are equal to
 zero at the  disconnected edges of GLs (at $x = -L$ in the upper GL and 
 at $x = L$ in the lower GL)

Fist from Eqs.~(21) and (22)  we obtain

\begin{equation}\label{eq25}
\delta\varphi_{+} + \delta\varphi_{-} = Ax,
\end{equation}
where $A$ is a constant.
Considering Eq.~(25), Eqs. (21) and (22) can be presented as

\begin{equation}\label{eq26}
\frac{d^2\delta\varphi_{+}}{dx^2} 
+\frac{2\omega(\omega + i\nu)}{s^2} \biggl(\delta\varphi_{+} - \frac{A}{2}x\biggr) = 0, 
\end{equation}
\begin{equation}\label{eq27}
\frac{d^2\delta\varphi_{-}}{dx^2} 
+\frac{2\omega(\omega + i\nu)}{s^2} \biggl(\delta\varphi_{-} - \frac{A}{2}x\biggr) = 0. 
\end{equation}

Solving Eqs.~(26) and (27) with boundary conditions (24) and (25),
we obtain

\begin{equation}\label{eq28}
\delta\varphi_{+} = \frac{\delta V_m}{2}\biggl(\frac{\displaystyle\frac{\cos\gamma_{\omega} x}
{\gamma_{\omega}\sin\gamma_{\omega} L} - x}
{\displaystyle\frac{\cos\gamma_{\omega} L}{\gamma_{\omega}\sin \gamma_{\omega} L} - L}\biggr),
\end{equation}

\begin{equation}\label{eq29}
\delta\varphi_{-} = -\frac{\delta V_m}{2}\biggl(\frac{\displaystyle\frac{\cos\gamma_{\omega} x}
{\gamma_{\omega}\sin\gamma_{\omega} L} + x}
{\displaystyle\frac{\cos\gamma_{\omega} L}{\gamma_{\omega}\sin \gamma_{\omega} L} - L}\biggr).
\end{equation}
Here
$\gamma_{\omega} = \sqrt{2\omega(\omega + i\nu)}/s$. Introducing the characteristic plasma frequency $\omega_p = \pi\,s/2\sqrt{2}L$, one obtains $\gamma_{\omega}L = \pi\sqrt{\omega(\omega + i\nu)}/2\omega_p$.

In the low modulation frequency limit 
(when $\omega \ll L^2/2s^2\nu = 4\omega_p^2/\pi^2\nu = \tau_M^{-1}$, where
$\tau_M^{-1}$ is the Maxwell relaxation time,
 and, hence, $|\gamma_{\omega}| L \ll 1$), one obtains $\delta\varphi_{+} = \delta V_m/2$
and $\delta\varphi_{-} = -\delta V_m/2$, i.e., the spatial distribution of the ac potential across the GLs is flat.

Figure~2 shows examples of the spatial distributions of the amplitudes of the ac potential across 
GLs, calculated for different modulation frequencies
using Eqs.~(28) and (29). As seen from Fig.~2, the amplitudes $|\delta\varphi_+|$
and $|\delta \varphi_{-}|$ are close to the amplitude of the applied ac voltage 
$\delta V_m/2$ at relatively low modulation frequencies ($\omega/2\pi = 0.20$ and 0.5~THz). However, with increasing  $\omega$, the amplitudes dramatically increase
(see the curves corresponding to $\omega/2\pi = 0.80$ and 1.00~THz).
This is attributed to the excitation of plasma oscillations whose amplitude grows
as the modulation frequency approaches to the plasma resonance frequency (see below).

\section{Resonant modulation}
Substituting $\delta\varphi_{+}$ and $\delta\varphi_{-}$ from Eqs.~(28) and (29) into Eq.~(21) and integrating over $dx$,
we obtain

$$
\delta m_{\omega} = \overline\beta_{\Omega}H
\frac{\displaystyle\exp\biggl[\frac{\hbar(\Omega - \Omega_0)}{2T}\biggr]}
{\biggl\{1 + \displaystyle\exp\biggl[\frac{\hbar(\Omega - \Omega_0)}{2T}\biggr]\biggr\}^2}
$$
\begin{equation}\label{eq30}
\frac{\sin(\gamma_{\omega}L)}{(\gamma_{\omega}L)[\cos(\gamma_{\omega}L) - (\gamma_{\omega}L)\sin(\gamma_{\omega}L)]}
\frac{\delta V_m}{\sqrt{2\overline{V}V_0}}.
 \end{equation}
 yielding  for the normalized modulation depth:
\begin{equation}\label{eq31}
\frac{\delta\,m_{\omega}}{\delta\,m_{0}}= 
\biggl|\frac{\sin (\gamma_{\omega}L)}
{(\gamma_{\omega}L)[\cos(\gamma_{\omega}L) - (\gamma_{\omega}L)\sin(\gamma_{\omega}L)]}\biggr|.
 \end{equation}
Using Eq. (31), the normalized modulation depth can also be expressed via the characteristic plasma frequency~$\omega_p$:

\begin{widetext}
\begin{equation}\label{eq31}
\frac{\delta m_{\omega}}{\delta m_{0}}= 
\biggl|
\frac{\displaystyle\biggl[\frac{2\omega_p}{\pi\sqrt{\omega(\omega + i\nu)}}\biggr]
\sin\biggl[\frac{\pi\sqrt{\omega(\omega + i\nu)}}{2\omega_p}\biggr]}
{\displaystyle
\biggl\{\cos\biggl[\frac{\pi\sqrt{\omega(\omega + i\nu)}}{2\omega_p}\biggr] - \biggl[\frac{\pi\sqrt{\omega(\omega + i\nu)}}{2\omega_p}\biggr]\sin\biggl[\frac{\pi\sqrt{\omega(\omega + i\nu)}}{2\omega_p}\biggr]\biggr\}}
\biggr|.
 \end{equation}

\end{widetext}%

\begin{figure}[t]\label{Fig.3}
\begin{center}
\includegraphics[width=6.3cm]{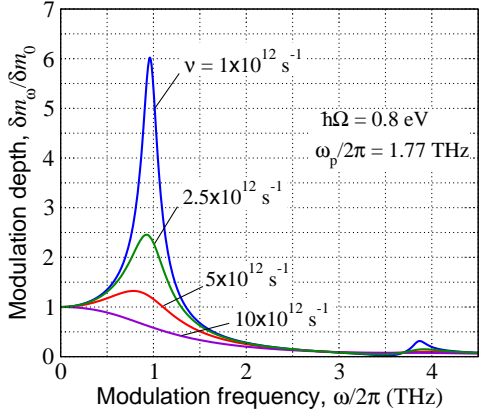}
\caption{Normalized modulation depth $\delta m_{\omega}/\delta m_0$
 versus modulation frequency $\omega/2\pi$
for different electron and hole collision frequencies $\nu$.
}
\end{center} 
\end{figure} 
\begin{figure}[t]\label{Fig.4}
\begin{center}
\includegraphics[width=6.3cm]{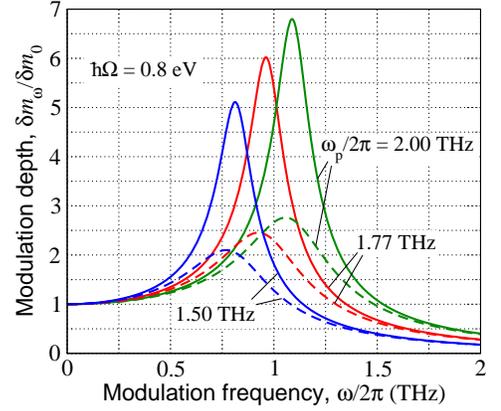}
\caption{
Normalized nodulation depth $\delta m_{\omega}/\delta m_0$ versus modulation frequency $\omega/2\pi$
for different plasma frequencies  $\omega_p$: solid lines
correspond to $\nu = 1\times10^{12}$~s$^{-1}$, dashed lines correspond to  $\nu = 2.5\times10^{12}$~s$^{-1}$
.}
\end{center} 
\end{figure}

At $\omega \ll \nu, \omega_p$,
Eq.~(32)  yields
 
\begin{equation}\label{eq33}
\frac{\delta m_{\omega}}{\delta m_{0}} \simeq
\biggl|\frac{\tan\sqrt{i\omega\tau_M}}
{\sqrt{i\omega\tau_M}}\biggr|.
 \end{equation}
Equations (19) and (33) provide the dependences of the modulation depth
on the structural parameters and the modulation frequency (particularly the frequency dependenceat the
 roll-off) obtained experimentally and  described in Ref.~{19}.

The character of the dependence of the modulation depth $\delta m_{\omega}$ on the modulation frequency $\omega$ depends on the collision frequency $\nu$. 
Figure 3 shows the frequency dependences (on $f = \omega/2\pi$)
of the normalized modulation depth $\delta m_{\omega}/\delta m_0$
calculated using Eq.~(32) for devices with different values of the collision frequency $\nu$. It is assumed that the plasma frequency $\omega_p/2\pi = 1.77$~THz.
For the characteristic plasma velocities $s = (3.75 - 7.5)\times 10^8$~cm/s
(as in the  estimate in the previous section), this frequency corresponds to $2L = 0.75 - 1.5~\mu$m,
i.e., to the length of GLs (size of the waveguide) close to the optical wavelength
under consideration.  

At relatively large collision frequencies $\nu$,
the modulation depth $\delta m _{\omega}$ monotonically decreases with increasing the modulation frequency [in line with Eq. (33)]. The curve for $\nu = 10\times 10^{12}$~s$^{-1}$ in Fig. 3 corresponds to $\tau_M \simeq 2\times 10^{-13}$~s
and the 3dB roll-off frequency $f_{3dB} \simeq 0.8$~THz.
In the devices with a smaller $\omega_p$, i.e., with a smaller  plasma-wave velocity,  $s$,  or larger length of GLs, $2L$, the Maxwell relaxation time is longer, so that
the roll-off frequency is smaller.
However, in the devices with relatively small $\nu \ll \omega_p$, 
$\delta m_{\omega}$  can exhibit a steep increase when $\omega$
approaches to the plasma resonance frequency.
The sharpness of the modulation depth peak and its height
rise with decreasing $\nu$ (i.e., with an increasing quality factor of the plasma resonances,
$Q \sim \omega_p/\nu$). This also seen in Fig. 4 (compare the solid and dashed lines). 

The plasma resonance frequencies, $\omega_n \propto \omega_p$. Here $n = 1, 2,3,...$ is the resonance index.  As can be derived from Eq. (32), the resonance frequencies are given by the solution of the following equation:

\begin{equation}\label{eq34}
\cot(\pi\omega_n/2\omega_p) = (\pi\omega_n/2\omega_p) .
 \end{equation}
Equation (34), in particular, yields $\pi\omega_1/2\omega_p \simeq 0.86 <  1$,
so that at $\omega_p/2\pi 1.77$~THz, one obtains  $\omega_1/2\pi \simeq 0.97$~THz.
The second (even) resonance corresponds to $\omega_2/2\pi \gtrsim 3.54$~THz. 
Thus, apart from a pronounced first resonance, a fairly weak second resonance is also seen in Fig.3.
A characteristic plasma frequency 
falls in the THz range.
As seen in Fig. 4, an increase in $\omega_p$ and, consequently, in  $\omega_n$ 
 shifts
 the positions of the peaks in the frequency dependence of the modulation depth. 
However, this shift can not be used for the electrical control because the bias voltage $V_0$ corresponds to the energy of photons of modulated radiation.
This is in contrast to  other THz voltage-controlled devices using the resonant excitation of plasma oscillations (see, for instance, Ref.~\cite{36}).

\section{Conclusions}

In summary,
we developed a device model for an  optical modulator based 
on  the double-GL structure  that was recently proposed and experimentally realized~\cite{19}.
The double-GL modulator
utilizes 
the variation of  absorption  due to  the
electrically controlled Pauli blocking effect. 
Our model accounts for the interband and intraband absorption and
the plasma effects in GLs determining the spatio-temporal distributions
of the electron and hole densities and the absorption coefficient,
The developed  model yields 
the dependence of the modulation depth on 
the control voltage for  strong but slow modulation signals
and for  small-signal modulation in a wide range of frequencies.
The dependence of the modulation depth
on the modulation frequency is determined by the relationship between
the collision frequency of electrons and holes and the characteristic plasma frequency (or between the latter and the Maxwell relaxation time).
At relatively large collision frequencies (or small plasma frequencies),
the modulation depth is a monotonically decreasing function of the modulation frequency. The obtained dependencies qualitatively explain the experimental results. However, we predict that in the double-GL structures with relatively weak disorder and, hence,
with a low collision frequencies, and a sufficiently high
 quality factor of the plasma oscillations,
the modulation depth exhibits a sharp maximum at the modulation frequency, which corresponds 
to the plasma resonance. The frequency of the latter falls in the THz range for typical parameter values.
 This opens up the possibility to use the double-GL structures for  
effective modulation of optical radiation by THz signals.

\section*{Acknowledgments}

This work was supported by the Japan Science and Technology Agency, CREST, 
The Japan Society for promotion of Science, Japan, and TERANO-NSF grant.
The work at RPI was supported by the US National Science Foundation and EAGER program
(monitored by Dr. Usha Varshne).


\begin{thebibliography}{99}







\bibitem{1}
A. H. Castro Neto, F. Guinea, N. M. R. Peres, K. S. Novoselov, and A. K. Geim,
Rev. Mod. Phys. {\bf 81}, 109 (2009). 
\bibitem{2}
J. M. Davlaty, S. Shivaraman, J. Strait, P. Geotge, M. Chandrashekhar, 
F. Rana, M. G. Sprncer, D. Veksler, and Y. Chen,
Appl. Phys. Lett.{\bf 93}, 131905 (2008).
\bibitem{3}
V.~Ryzhii, M.~Ryzhii, and T.~Otsuji,
J. Appl. Phys. {\bf 101}, 083114 (2007).
\bibitem{4}
M.Ryzhii and V.Ryzhii,
Jpn. J. Appl. Phys. (Express Lett.) {\bf 46,} L151 (2007). 

\bibitem{5}
F.~Rana, IEEE Trans. Nanotechnol. {\bf 7}, 91 (2008).
\bibitem{6}

V.Ryzhii, M.Ryzhii, and T.Otsuji,
Phys. Stat. Sol. (c) {\bf 5}, 261 (2008).



 \bibitem{7}
A.~Satou, F.~T.~Vasko, and V.~Ryzhii,
Phys. Rev. B {\bf 78}, 115431 (2008).


\bibitem{8} 
A.~A.~Dubinov, V.~Ya. Aleshkin, M.~Ryzhii, T.~Otsuji,
and V.~Ryzhii,
Appl. Phys. Express {\bf 2}, 092301  (2009). 

\bibitem{9}
V.~Ryzhii, M.~Ryzhii, A.~Satou, T.~Otsuji, A.~A.~Dubinov, and V.~Ya. Aleshkin,
J. Appl. Phys. {\bf 106}, 084507 (2009) 
 
\bibitem{10}
V.~Ryzhii,  A.~A.~Dubinov, T. Otsuji,V. Mitin, and M.S.Shur,
%
J. Appl. Phys. {\bf 107}, 054505 (2010). 
 
\bibitem{11} 
V.~Ryzhii, V.~Mitin, M.~Ryzhii, N.~Ryabova, and T.~Otsuji,
Appl. Phys. Express {\bf 1}, 063002 (2008).


\bibitem{12}
V. Ryzhii and M. Ryzhii,
Phys. Rev. B {\bf 79}, 245311 (2009).  
\bibitem{13}
F.~Xia, T.~Mueller, Y-M.~Lin, A.~Valdes-Garsia, and F.~Avouris,
Nature Nanotecnology, {\bf 4}, 839 (2009).

\bibitem{14}
T.~Mueller, F. Xia, and  and F.~Avouris,
Nature Photon., {\bf 4}, 297 (2010).


\bibitem{15}
X.D.Xu, N.M.Gabor, J.S.Allen, A.M. van der Zande, and P.L.McEuen,
Nano Lett., {\bf 10}, 562 (2010).


\bibitem{16}
F. Bonaccorso, Z. Sun, T. Hasa, and A.C. Ferrari,
Nat. Photonics, {\bf 4}, 611 (2010).



\bibitem{17}
P.~Neugebauer, M.~Orlita, C.~Faugeras,  A.-L.~Barra, and M.~Potemski,
Phys. Rev. Lett. {\bf 103}, 136403 (2009).
%
\bibitem{18}
 M.~Orlita and M.~Potemski,
Semicond. Sci. Technol. {\bf 25}, 063001 (2010).
%
\bibitem{19}
M. Liu, X. Yin, and X. Zhang,
Nano Lett. {\bf 12}, 1482 (2012)








\bibitem{20}
S.~J.~Allen, Jr., D.~C.~Tsui, and R.~A.~Logan,
Phys. Rev. Lett. {\bf 38}, 980 (1977).

\bibitem{21}
D.~C.~Tsui, E.~Gornik, and  R.~A.~Logan,
Solid State Commun. {\bf 35}, 875 (1980).

 \bibitem{22}
M. Dyakonov and M. Shur,
IEEE Trans. Electron Devices {\bf 43}, 1640  (1996).

\bibitem{23}
W.~Knap, Y.~Deng, S.~Rumyantsev, J.-Q.~Lu, M.~S.~Shur, C.~A.~Saylor,
and L.~C.~Brunel,
Appl.~Phys.~Lett. {\bf 80}, 3433  (2002). 

\bibitem{24}
X.~G.~Peralta, S.~J.~Allen, M.~C.~Wanke, N.~E.~Harff, J.~A.~Simmons,
M.~P.~Lilly, J.~L.~Reno, P.~J.~Burke, and J.~P.~Eisenstein,
Appl.~Phys.~Lett. {\bf 81}, 1627  (2002).

\bibitem{25}
T. Otsuji, M. Hanabe and O.~Ogawara,
Appl. Phys. Lett. {\bf 85}, 2119  (2004).


\bibitem{26}
J.~Lusakowski, W.~Knap, N.~Dyakonova, L.~Varani, J.~Mateos, T.~Gonzales,
Y.~Roelens, S.~Bullaert, A.~Cappy and K.~Karpierz,
J.~Appl. Phys. {\bf 97},   064307 (2005). 

\bibitem{27}
F.~Teppe, W.~Knap, D.~Veksler, M.~S.~Shur, A.~P.~Dmitriev,
V.~Yu.~Kacharovskii, and S.~Rumyantsev,
Appl. Phys. Lett. {\bf 87}, 052105 (2005).


\bibitem{28}
V.Ryzhii, A.Satou, W.Knap, and M.S.Shur.
J.~Appl. Phys. {\bf 99},  084507  (2006).

 \bibitem{29}
A.~El Fatimy, F.~Teppe, N.~Dyakonova,  W.~Knap,
D.~Seliuta, G.~Valusis, A.~Shcherepetov, Y. Roelens, S.~Bollaert, A.~Cappy,
and S.~Rumyantsev,
Appl. Phys. Lett. {\bf 89}, 131926 (2006).

\bibitem{30}
J.~Torres, P.~Nouvel, A.~Akwaoue-Ondo, L.~Chusseau, F.~Teppe,  
A.~Shcherepetov, and S.~Bollaert, Appl. Phys. Lett.{\bf 89}, 201101 (2006)

\bibitem{31}
V. Ryzhii,
Jpn. J. Appl. Phys. {\bf 45}, L923 (2006).


\bibitem{32}O. Vafek, Phys. Rev. Lett. {\bf 97},266406 (2006).

\bibitem{33}
L.~A.~Falkovsky and A.~A.~Varlamov, Eur. Phys. J. B {\bf 56}, 281 (2007)
\bibitem{34}
V. Ryzhii, A.Satou, and T. Otsuji,
J.~Appl. Phys. {\bf 101 },024509    (2007).


\bibitem{35}
D. Svintsov, V. Vyurkov, S. Yurchenko, T. Otsuji, and V. Ryzhii,
J. Appl. Phys. {\bf 111}, 083715 (2012).

\bibitem{36}
V. Ryzhii, T. Otsuji, M. Ryzhii, and M. S.  Shur,
J. Phys. D: Appl. Phys. {\bf 45}, 302001 (2012).













\end{thebibliography}
\end{document}